\documentclass[11pt]{article}
\usepackage{moriond,epsfig}
\usepackage{epstopdf}

\def\Journal#1#2#3#4{{#1} {\bf #2}, #3 (#4)}
\def\Journall#1#2#3#4{{#1} {\bf #2}, (#3) #4}

\def\NPB{{\em Nucl. Phys.} B}
\def\PLB{{\em Phys. Lett.}  B}

\def\PRD{{\em Phys. Rev.} D}

\def\JINST{{\em JINST}}
\def\EPJ{{\em EPJ} C}
\def\JHEP{{\em JHEP}}

\def\be{\begin{equation}}
\def\ee{\end{equation}}
\def\bea{\begin{eqnarray}}
\def\eea{\end{eqnarray}}

\def\ZZ{ZZ\rightarrow 4 \ell}
\def\ZZnu{ZZ\rightarrow \ell\ell\nu\nu}
\def\WZ{WZ\rightarrow \ell\nu\ell\ell}
\def\WW{WW\rightarrow \ell\nu\ell\nu}
\def\WZsem{WW+WZ \rightarrow \ell\nu jj}
\def\Zg{Z\gamma \rightarrow \ell\ell\gamma}
\def\Zgn{Z\gamma \rightarrow \nu\nu\gamma}
\def\Wg{W\gamma \rightarrow \ell\nu\gamma}
\def\ifb{fb^{-1}}

\newcommand{\Photo}{}

\begin{document}

\vspace*{4cm}
\title{DIBOSON PRODUCTION CROSS SECTION AT LHC}

\author{ VINCENZO P. LOMBARDO \\ (on behalf of the ATLAS and CMS Collaborations)}
\address{Laboratoire d'Annecy-le-Vieux de Physique des Particules\\
9 Chemin de Bellevue, Annecy-le-Vieux, France}

\maketitle\abstracts{
This paper presents an overview of the diboson production cross-section measurements
and constraints on anomalous triple-gauge boson couplings performed by the ATLAS and 
CMS collaborations using proton-proton collisions produced at a centre-of-mass energy
of $\sqrt{s}$ = 7 and 8 TeV at LHC. Results for all combinations of $W, Z$ and $\gamma$ gauge 
bosons (excluding $\gamma\gamma$) are presented with emphasis on the new $WZ$ and $ZZ$ 
production cross sections measured by ATLAS at $\sqrt{s}$ = 8 TeV and on the new constraints 
on anomalous triple-gauge couplings set by CMS in the $WW$ and $Z\gamma$ modes.
} 

\section{Introduction}
In the Standard Model (SM) of particle physics, measurements of diboson final states at 
the TeV scale provide excellent tests of the electroweak sector. Any deviation of the 
diboson production cross sections or kinematic distributions from the SM predictions 
may be an indication of anomalous triple-gauge boson couplings (aTGC) and of the existence 
of new particles 
such as those predicted by technicolor models, little Higgs and Randall-Sundrum graviton models,
to mention only a few examples. It is therefore very important to have precise diboson measurements 
as well as reliable and accurate theoretical predictions for these processes. In addition 
non-resonant diboson production processes must be understood in detail as they represent a 
significant background to the measurements 
of the Higgs boson. The diboson results presented in this note are based on the proton-proton 
collision data collected by the ATLAS~\cite{atlas} and CMS~\cite{cms} detectors at the LHC~\cite{lhc} 
in 2011 and 2012. 

\section{Diboson production cross-section measurements} 

The measurements of the total production cross sections of $pp \rightarrow W\gamma + X$ 
and $pp \rightarrow Z\gamma + X$ have been performed in the decay channels $\Wg$ and $\Zg$~\cite{wzgamma_atlas,wzgamma_cms} 
by requiring a $W$ or a $Z$ boson candidate in association with an isolated photon. The 
photon is required to be separated by the 
lepton to suppress QED Final State Radiation. The dominant background includes $W/Z + jets$ 
and $\gamma + jets$ where jet-induced photons or jet-faking leptons are selected. In addition 
to the total cross section both the ATLAS and CMS Collaborations have unfolded the 
$E_T^{\gamma}$ spectrum observing a reasonable agreement with theoretical predictions. 
The $\Zgn$ mode has also been measured by the two Collaborations~\cite{wzgamma_atlas,zgamma_cms} 
by requiring a tighter selection on the photon and on the missing transverse energy ($E_T^{miss}$) 
in order to  suppress the backgrounds. A fair agreement with theoretical predictions is observed.

The $pp \rightarrow WZ + X$ production cross section has been measured in the $\WZ$ channel
by both ATLAS and CMS at a centre-of-mass energy $\sqrt{s}$ = 7 TeV with 4.6 and 1.1 $\ifb$, 
respectively~\cite{wz_atlas,wz_cms}. The signal region is selected by requiring three isolated 
high-$p_T$ leptons and $E^{miss}_T$. A tighter identification requirement on the lepton coming from 
the W decay reduces the Z+jet background where jet-faking leptons are selected. In addition 
to the the total cross section ATLAS has measured the inclusive $WZ$ cross section as function 
of the transverse momentum of the $Z$ boson and of the invariant mass of the $WZ$ system.
Recently, ATLAS has also released the first measurement~\cite{wz_atlas8} of the $pp \rightarrow WZ +X$ 
production cross section at 8 TeV with an integrated luminosity of 13 $\ifb$. The measured 
cross section is $20.3^{+0.8}_{-0.7}(stat)^{+1.2}_{-1.1}(syst)^{+0.7}_{-0.6}(lumi)$ pb, in 
excellent agreement with the next-to-leading order (NLO) QCD predictions provided by MCFM + 
CT10 parton distribution functions~\cite{mcfm,ct10}.    

The $pp \rightarrow WW + X$ production cross section has been measured by ATLAS and CMS 
at 7 TeV with 4.6 and 4.9 $\ifb$, respectively~\cite{ww_atlas,ww_cms}. 
The selection 
of the $\WW$ channel requires two opposite charged isolated leptons. The Drell-Yan background 
in the $ee$ and $\mu\mu$ channels is suppressed by means of a Z veto and a requirement on 
a $E_T^{miss}$-related variable. In addition a jet veto is applied to reduce the 
top-quark background. The cross section measured by ATLAS is $51.9\pm2.0(stat)\pm3.9(syst)\pm2.0(lumi)$
pb and the one measured by CMS is $52.4\pm2.0(stat)\pm4.5(syst)\pm1.2(lumi)$ pb. They agree to 
each other but they are higher than the NLO calculations provided by MCFM+CT10 which predicts 
$44.7^{+2.1}_{-1.9}$ pb. The CMS collaboration has also measured the $WW$ cross section at 8 TeV 
with a dataset corresponding to an integrated luminosity of 3.5 $\ifb$~\cite{ww_cms8}. The measured cross
section is $69.9\pm2.8(stat)\pm5.6(syst)\pm3.1(lumi)$ pb confirming a higher cross section with 
respect to the NLO predictions calculated by MCFM+CT10 which gives $57.3^{+2.4}_{-1.6}$ pb.

Both ATLAS and CMS have also measured the total cross section of $\WZsem$ at 7 TeV with 
4.7 and 5.0 $\ifb$, respectively~\cite{ww_wz_atlas,ww_wz_cms}. The request of exactly 
two high-$p_T$ jets in addition  to an isolated lepton and $E_T^{miss}$ is used 
to select the signal region. After these selection requirements the $W/Z + jets$ background 
is still huge compared to signal and a binned maximum-likelihood fit to the dijet invariant mass 
has been used to extract the signal yield and measure the cross section. Both ATLAS and CMS report 
results which are in good agreement with the SM predictions.

\begin{figure}[htb] \centering
  \begin{minipage}{17pc}
    \includegraphics[width=18pc]{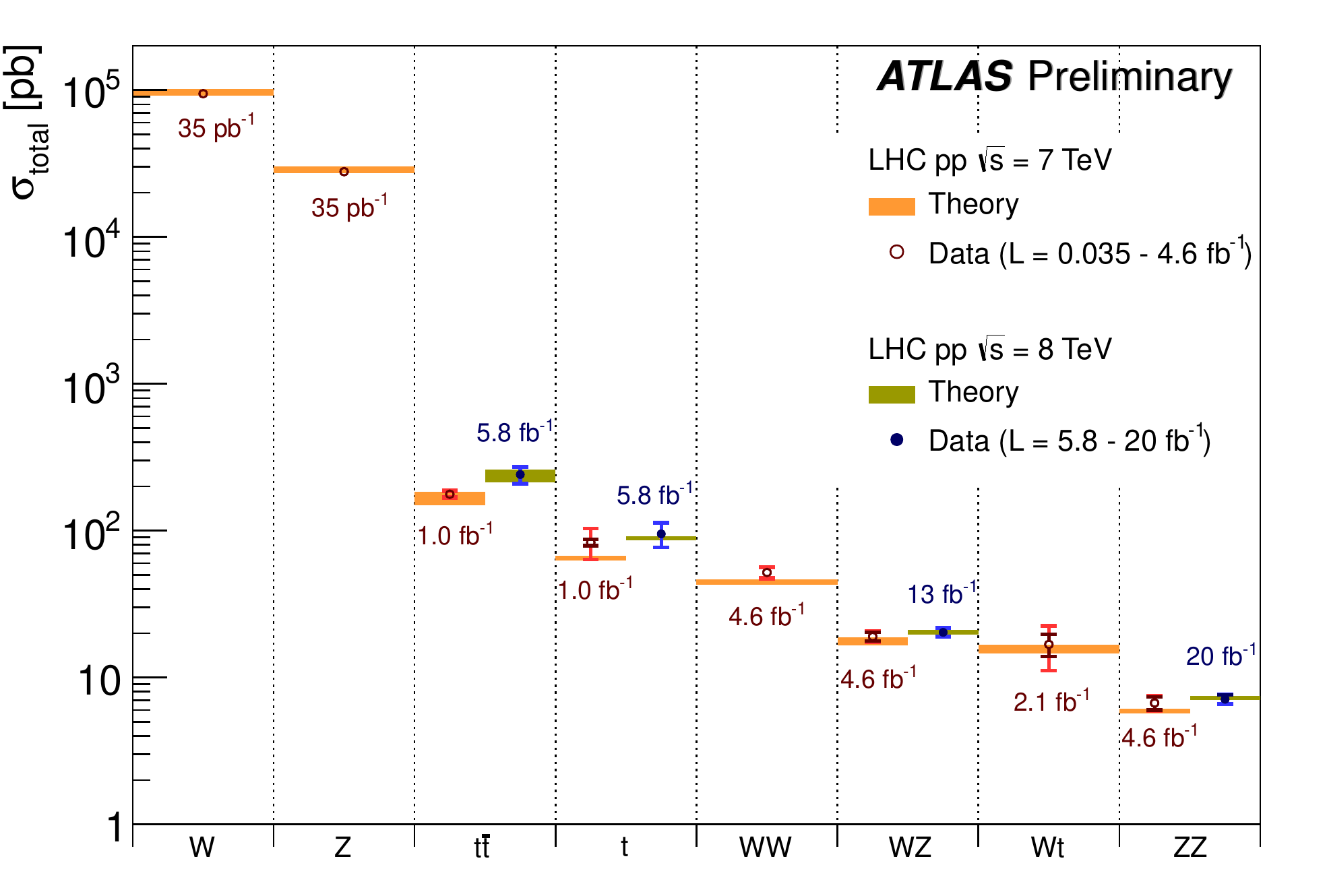}
    \caption{\label{fig:fig1} Summary plot of ATLAS cross-section measurements~\protect\cite{atlastwiki}.}
  \end{minipage}\hspace{2pc}
  \begin{minipage}{17pc}
    \includegraphics[width=19pc]{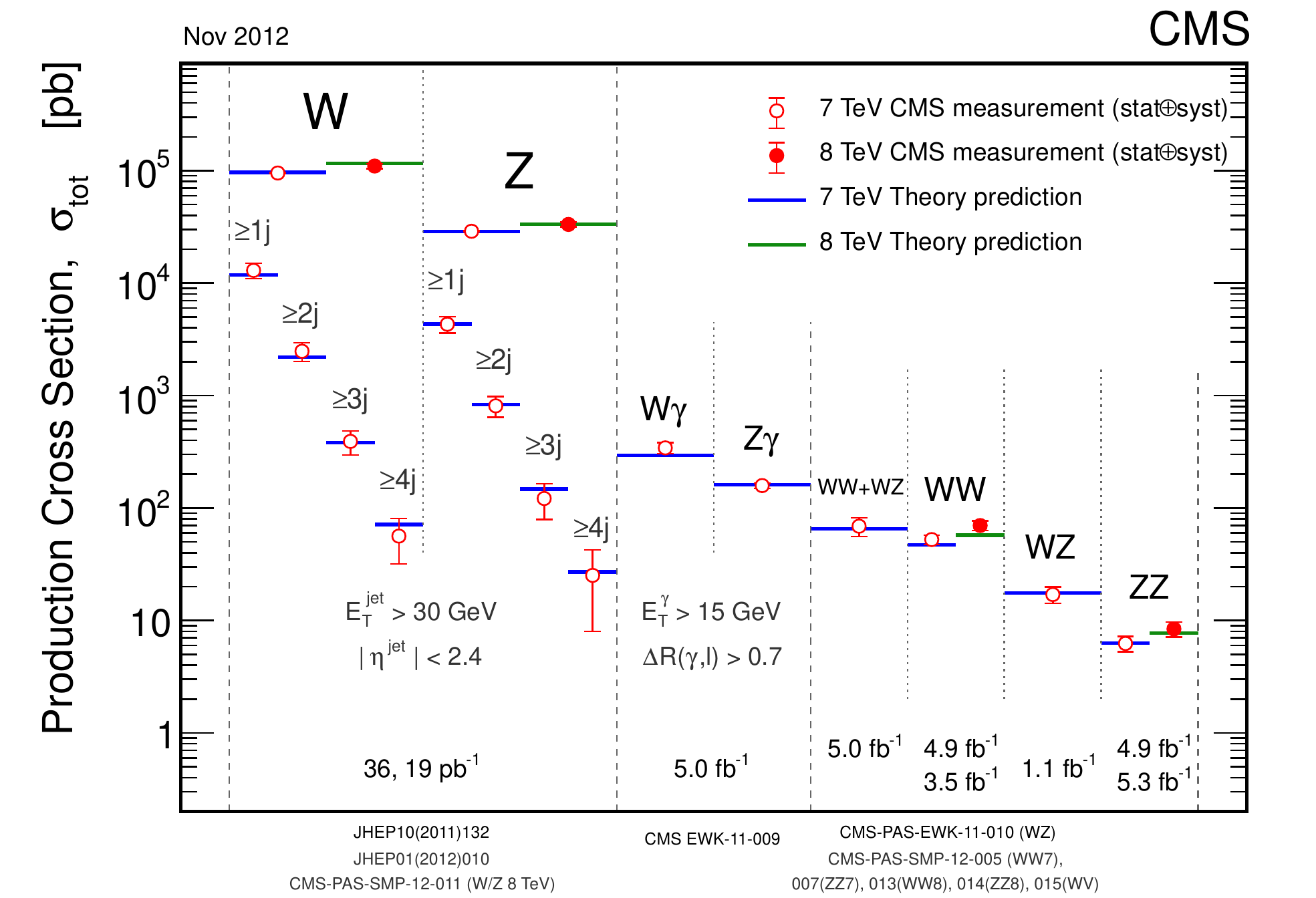}
    \caption{\label{fig:fig2} Summary plot of CMS cross-section measurements~\protect\cite{cmstwiki}.}
\end{minipage} \end{figure}

The $pp \rightarrow ZZ + X$ production cross section has been measured by ATLAS and CMS at
the centre-of-mass energy of $\sqrt{s}$ = 7 and 8 TeV~\cite{zz_atlas,zz_cms,ww_cms8}. 
The $\ZZ$ final state requires two pairs of isolated charged leptons (e or $\mu$) which are 
compatible with two on-shell Z bosons. The ATLAS Collaboration has also measured the $ZZ$ cross 
section at $\sqrt{s}$ = 7 TeV in the $\ZZnu$ channel~\cite{zz_atlas} by applying a tight cut 
on a $E^{miss}_T$-related variable in order to suppress the dominant $Z + jets$ background. 
On the other hand, the CMS measurement at 8 TeV includes the $ZZ \rightarrow 2\ell 2\tau$ mode. 
The ATLAS Collaboration has recently released an updated measurement of the $\ZZ$ cross section 
with 20 $\ifb$ at $\sqrt{s}$ = 8 TeV~\cite{zz_atlas8}. The measured cross section is 
$7.1^{+0.5}_{-0.4}(stat)\pm 0.3(syst)\pm 0.2(lumi)$ pb which agrees very  well with the NLO 
calculation provided by MCFM with CT10 predicting $7.2^{+0.3}_{-0.2}$ pb. 

A summary of the up-to-date diboson production cross-section measurements
performed by ATLAS and CMS are shown in Figure~\ref{fig:fig1} and ~\ref{fig:fig2}. The diboson
cross sections, as well as several other SM production cross sections, are compared to their 
theoretical predictions.

\section{Anomalous triple-gauge boson couplings}
In the SM, triple-gauge boson couplings (TGC) are completely fixed by the gauge structure of the electroweak theory 
and any deviation from the SM couplings indicates new physics beyond the SM. An effective lagrangian featuring such 
anomalous triple-gauge boson couplings (aTGCs) can be constructed and then compared to the experimental data. 
Under some general assumptions, aTGCs can be parameterized using the set of parameters shown 
in Table~\ref{tab:tab1}~\cite{hagiwara,ellison}. All these parameters are equal to zero in the SM. Anomalous couplings are expected 
to modify the total production cross section as well as the kinematic distributions of the diboson processes, in particular 
in high momentum or high mass regions. The ATLAS and CMS Collaborations have searched for anomalous triple-gauge boson 
couplings and found no deviation from the SM values. Figure~\ref{fig:fig5},~\ref{fig:fig6},~\ref{fig:fig7} and~\ref{fig:fig8} 
show the limits at $95\%$ confidence level on the aTGCs set by ATLAS and CMS as well as those set by LEP and Tevatron.
  
\begin{table}[htb]
\begin{center}
\begin{tabular}{|c|c|c|}
\hline
coupling & parameters & channel \\ 
\hline
$WW\gamma$ & $\lambda_{\gamma}, \Delta k_{\gamma}$& $WW, W\gamma$   \\
$WWZ$ & $\lambda_{Z}, \Delta k_{Z}$, $\Delta g_1^Z$ & $WW,WZ$ \\
$ZZ\gamma$ & $h_3^Z, h_4^Z$& $Z\gamma$\\
$Z\gamma\gamma$ & $h_3^{\gamma}, h_4^{\gamma}$ & $Z\gamma$\\
$Z\gamma Z$ & $f_{40}^{Z}, f_{50}^{Z}$& $ZZ$\\
$ZZZ$ & $f_{40}^{\gamma}, f_{50}^{\gamma}$& $ZZ$\\
\hline
\end{tabular}
\caption{\label{tab:tab1} List of TGC parameters commonly used in ATLAS and CMS~\protect\cite{hagiwara,ellison}.} 
\end{center}
\end{table}
\noindent
The CMS Collaboration has recently released new limits on aTGCs in the $\WW$ channel analyzing 4.9 $\ifb$ 
of data collected at a centre-of-mass energy of $\sqrt{s}$ = 7 TeV~\cite{ww_cms}. These limits are 
shown in Figure~\ref{fig:fig5} and ~\ref{fig:fig6}, together with previous results. In addition 
CMS has extracted limits on aTGCs in the $\Zgn$ channel~\cite{zgamma_cms} setting tighter constraints on 
aTGCs with respect to those extracted in the $\Zg$ analyses~\cite{wzgamma_cms,wzgamma_atlas}.  

\begin{figure}[htb] \centering
  \begin{minipage}{16pc}
    \includegraphics[width=17pc]{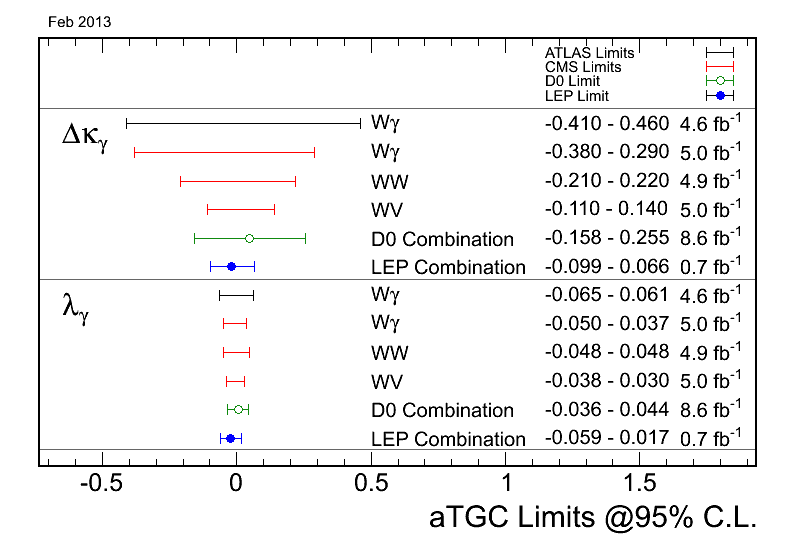}
    \caption{\label{fig:fig5} Limits on $WW\gamma$ anomalous triple-gauge couplings~\protect\cite{atgc} .}
  \end{minipage}\hspace{3pc}
  \begin{minipage}{16pc}
    \includegraphics[width=17pc]{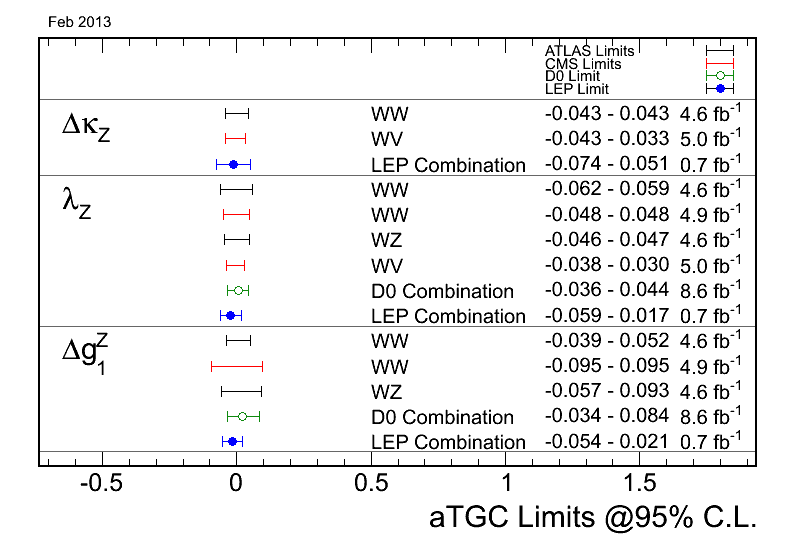}
    \caption{\label{fig:fig6} Limits on WWZ anomalous triple-gauge couplings~\protect\cite{atgc}.}
\end{minipage} \end{figure}

\begin{figure}[htb] 
\hspace{-1.8pc}
  \begin{minipage}{16pc}
    \includegraphics[width=18.5pc]{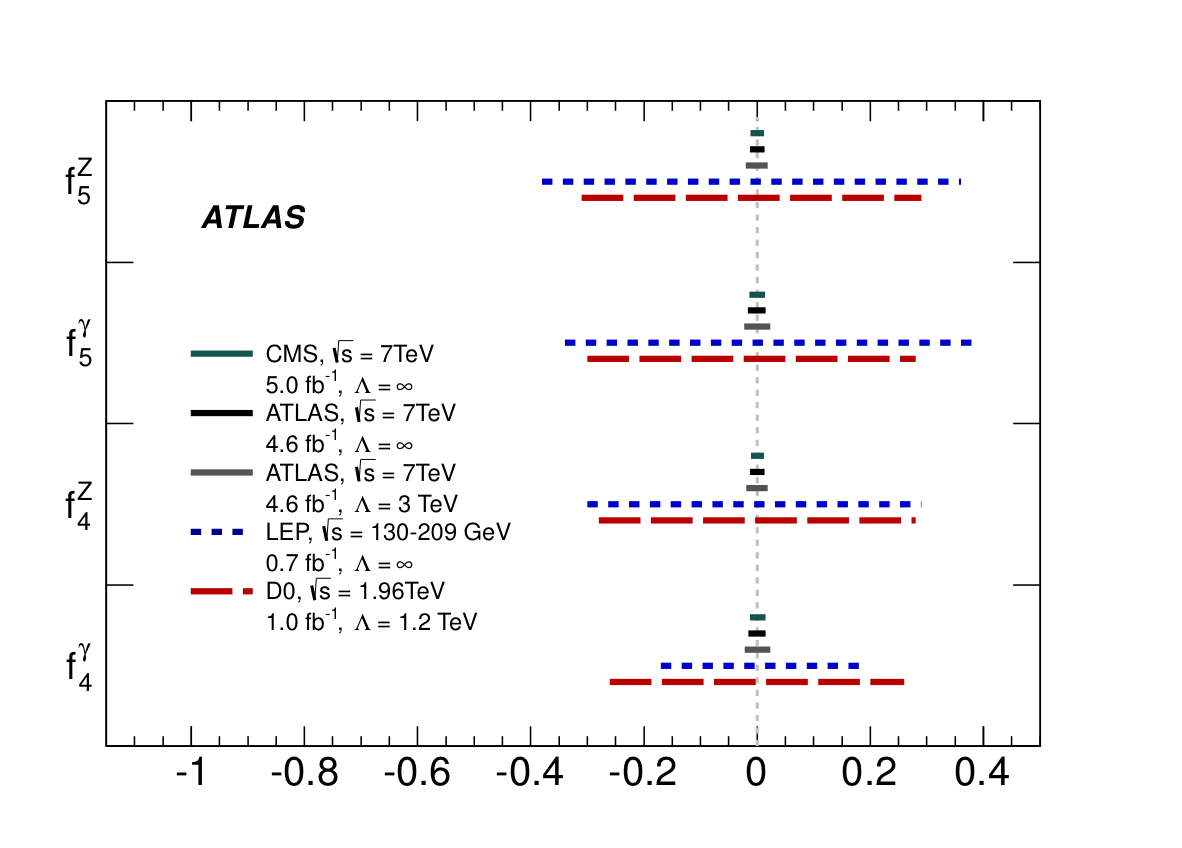}
    \caption{\label{fig:fig7} Limits on $Z\gamma Z$ and $ZZZ$ anomalous triple-gauge couplings~\protect\cite{zz_atlas}.}
  \end{minipage}\hspace{0.4pc}
  \begin{minipage}{16pc}
    \vspace{1pc}
    \includegraphics[width=24pc]{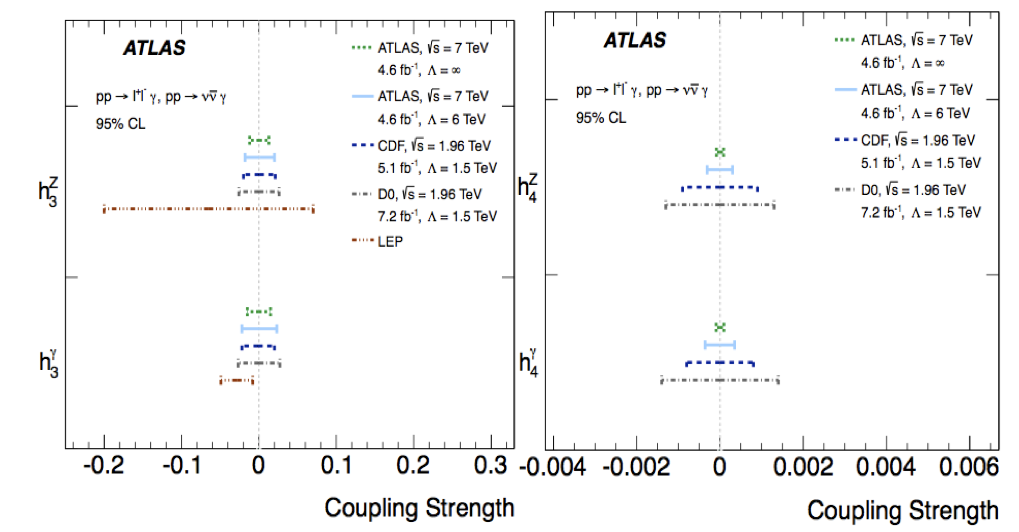}
    \caption{\label{fig:fig8} Limits on $ZZ\gamma$ and $Z\gamma\gamma$ anomalous triple-gauge couplings~\protect\cite{wzgamma_atlas}.}
\end{minipage} \end{figure}

\section{Summary}
Measurements of the production cross sections of diboson final states performed by the ATLAS and CMS 
Collaborations at a centre-of-mass energy of $\sqrt{s}$ = 7 and 8 TeV have been presented. The total 
production diboson cross sections are in reasonable agreement with the SM predictions within the 
uncertainties. Several distributions have also been unfolded and directly compared with the available 
theoretical distributions. A search for anomalous triple-gauge couplings have been performed by both 
ATLAS and CMS with almost 5 $\ifb$ of data collected at 7 TeV at LHC. Limits on aTGGs are set in all 
channels showing no deviation from the SM couplings.

\end{document}



